# Visualizing Chiral Edge Modes in Twisted Cuprate Superconductors via Scanning-Probe Quantum Sensing

Senlei Li[1,+], Xiaomeng Cui[2,+], Renzhi Sun[1], Lingjie Zhou[1], Yiran Zhao[1], Kenji Watanabe[3], Takashi Taniguchi[4], Genda Gu[5], Hailong Wang[1], Kartiek Agarwal[6,*], Philip Kim[2,*] and Chunhui Rita Du[1,*]

[1]School of Physics, Georgia Institute of Technology, Atlanta, Georgia 30332, USA
[2]Department of Physics, Harvard University, Cambridge, Massachusetts 02138, USA
[3]Research Center for Electronic and Optical Materials, National Institute for Materials Science, Tsukuba 305-0044, Japan
[4]Research Center for Materials Nanoarchitectonics, National Institute for Materials Science, Tsukuba 305-0044, Japan
[5]Department of Condensed Matter Physics and Materials Science, Brookhaven National Laboratory, Upton, New York 11973, USA
[6]Material Science Division, Argonne National Laboratory, Lemont, Illinois 60548, USA

[*]Corresponding authors: kagarwal@anl.gov, pkim@physics.harvard.edu, and cdu71@gatech.edu
[+]These authors contributed equally.

**Abstract**: Recently, unconventional superconductivity hosted by twisted van der Waals (vdW) heterostructures has received immense interest due to the exotic pairing symmetry, electronic interactions and nontrivial topological nature that are naturally relevant to the fast-advancing quantum technologies. Here, we report scanning-probe quantum sensing of nanoscale electromagnetic behaviors of twisted vdW cuprate superconductors. Using single-spin relaxometry, we directly visualize edge modes spontaneously formed in twisted $Bi_2Sr_2CaCu_2O_{8+x}$ (BSCCO). By investigating temperature dependent variations of edge-state-induced quantum spin relaxation, we experimentally evaluate the magnitude of nontrivial topological band gap opened at nodal points in twisted BSCCO and its critical temperature behaviors under different twist angles. We further observe alternating chiral domains defined by edge modes, exploring experimental signatures of in-plane magnetic field-induced topology in 45° twisted BSCCO. Our results advance the current understanding of twisted vdW nodal superconductors, presenting an appealing high-temperature topological superconducting material platform for cutting-edge quantum innovation.

Twist engineering of two-dimensional (2D) van der Waals (vdW) heterostructures has opened an emerging topic of moiré quantum matter at the forefront of condensed matter physics, materials science and electronics research(*1–4*). By stacking layers of atomically thin vdW crystals with a desirable twist angle or a lattice mismatch, a plethora of exotic electronic, photonic and magnetic phases can be created thanks to the tunable lattice interactions that are unavailable in naturally formed bulk and 2D materials(*3*, *5*). Notable examples include twisted/moiré superconductors(*1*, *2*, *6–10*), twisted magnetism(*11–15*), moiré excitons(*16–18*), and correlated topological electronic states(*2*, *19–21*).

Layered cuprate superconductor $Bi_2Sr_2CaCu_2O_{8+x}$ (BSCCO) recently emerged as a new member in the waves of twisted vdW quantum material research(*6*, *22–24*). Pristine BSCCO is a nodal $d$-wave superconductor down to the monolayer thickness limit(*22*, *25*, *26*). Theoretical work predicts that twist engineering can modify and induce topology to the band structure of BSCCO(*7*, *27*). The nodal gap signature vanishes in the ideal 45° twist case, and the corresponding bulk state becomes fully gapped with spontaneous formation of topologically protected chiral edge modes(*7*). The fascinating prospect of high-temperature topological superconductivity has brought immense research interest to twisted BSCCO due to its potential as a novel material platform for developing fault-tolerant quantum computing(*7*, *28*). Despite the intriguing fundamental physics and enormous technological promise, the ongoing research on twisted BSCCO remains in its infancy. In most of the previous studies, knowledge on topology, band structures, and symmetry analysis of twisted BSCCO were mainly based on microscopic modelling or inferred from electrical transport measurements(*6*, *7*, *29–33*). Direct observation of edge modes, chiral edge currents(*27*, *34*, *35*), and experimental evaluation of the topological bandgap in twisted BSCCO remain elusive in the current state of the art.

Here, we report scanning-probe quantum sensing of nanoscale electromagnetic response of twisted BSCCO. Taking advantage of quantum spin relaxometry(*15*, *36–39*), we directly visualize gapless edge modes in 45° twisted BSCCO. By investigating temperature dependent variations of edge states, we present an experimental approach capable of evaluating the magnitude of the topological bandgap opened in nodal points of twisted BSCCO. Our results show that observed edge states exist over a narrow window of twist angle in the vicinity of 45° and feature a characteristic twist angle dependent topological transition temperature ($\tilde{T}_c$) that is different from the superconducting transition point ($T_c$). We further explore magnetic field-induced topology and observe signatures of alternating chiral $d \pm id'$ domains defined by chiral edge modes in 45° twisted BSCCO.

## Scanning-probe quantum sensing platform

We first review the pertinent material properties and experimental platform as illustrated in Fig. 1A. Twisted BSCCO devices in the current study are prepared by a cryogenic, solvent-free vdW transfer technique(*6*). Figure 1B shows a microscopy image of a prepared 45° twisted BSCCO device. Thickness of the top and bottom BSCCO layers is ~10 and ~30 nm, respectively, and the target twist angle $\theta$ can be controlled with an accuracy of $\pm 0.1°$ in stacking processes [see section S1 of (*40*) for details]. As a $d$-wave superconductor, the superconducting order parameter of BSCCO has a $d_{x^2-y^2}$ symmetry with gapless density of states along the nodal directions. The 45° twisted top layer exhibits $d_{xy}$ order parameter in the untwisted coordinate frame. Cooper pair tunneling between the top and bottom BSCCO layers results in a hybrid formation of $d_{x^2-y^2} \pm id_{xy}$ in 45° twisted BSCCO as shown in Fig. 1C(*6*, *7*).

We utilize an nitrogen-vacancy (NV) center implanted at the end of a diamond nanopillar(*12*, *15*, *37*, *41–46*) to perform scanning-probe quantum sensing of twisted BSCCO devices. NV sensing takes advantage of quantum mechanical nature of an $S = 1$ electron spin to achieve sensitive detection of weak magnetic fields(*12*, *15*, *41*, *43*, *47–54*). The ultimate spatial resolution of our measurements is determined by the vertical NV-to-sample distance, which is ~60 nm in the current study(*15*). We first introduce static NV magnetometry to investigate superconducting properties of the prepared 45° twisted BSCCO device. Figures 1D and 1E present scanning NV imaging of static magnetic field maps of a surveyed sample area covering both pristine and twisted BSCCO at 88 K and 2 K. A small out-of-plane (OOP) external magnetic field $B_z \sim 10$ G is applied in these measurements for optimal sensing performance. We use the linear Zeeman effect of the NV center to detect local static magnetic field component along the NV spin direction(*12*, *15*). Magnitude of the magnetic field $B_s$ emanating from the BSCCO sample can be deduced from field-induced splitting of NV spin energies using optically detected magnetic resonance measurements [see section S2 of (*40*) for details]. Note that $B_z$ has been subtracted in our presentations to highlight the intrinsic field contribution of the BSCCO sample.

One can see that the BSCCO device generates a vanishingly small static magnetic field in both twisted and untwisted sample areas when above the superconducting transition temperature $T_c$ (Fig. 1D). As temperature falls below $T_c$, Meissner effect-induced magnetic flux expulsion modifies the local magnetic field environment of the sample(*41*), producing negative $B_s$ in the bulk areas of the BSCCO flake (Fig. 1E). Note that the sign of $B_s$ represents its orientation relative to the external magnetic field ($B_z$). The weaker magnetic field observed in the top BSCCO flake area is attributed to suppressed supercurrents hosted in a thinner superconductor. Figure 1F summarizes the temperature dependent evolution of magnetic field $|B_s|$ measured at three representative sample sites located in the top, bottom, and twisted BSCCO areas. $|B_s|$ exhibits a continuous decay all the way down to zero, featuring a characteristic superconducting phase transition in BSCCO. $T_c$ is determined to be ~86 K, which is close to the bulk value(*6*, *22*).

**Visualizing topological edge modes in 45° twisted BSCCO**

After examining the superconducting properties, we now demonstrate NV imaging of topological edge modes in 45° twisted BSCCO. We start with the discussion about topological band structure of twisted BSCCO as shown in Fig. 2A. Due to hybridization of the order parameter, a small bandgap is opened at local nodal points in 45° twisted BSCCO, and two gapless chiral edge states carrying quasiparticle currents with superposition of electrons and holes are spontaneously formed in the top and bottom BSCCO layers (Fig. 2B)(*7*, *34*). Bogoliubov quasiparticles are thermally populated in the edge states and potentially extending to the fully gapped bulk energy spectrum (depending on the temperature energy scale) with an occupation function following the Fermi-Dirac distribution. When temperature is (well) below $T_c$, magnetic noise emanating from 2D superconducting BSCCO is largely described by quasiparticle scattering(*55*), which dominates over longitudinal current fluctuations and potential spin fluctuations. Our quantum sensing measurements utilize the interaction between quasiparticle current fluctuations in twisted BSCCO and the NV sensor. Through random scattering processes, a quasiparticle with a frequency $f + \Delta f$ can be scattered to another state with a frequency $f$, leading to emission of fluctuating magnetic fields at a frequency $\Delta f$. When $\Delta f$ matches the NV electron spin resonance (ESR) frequency $f_{\mathrm{ESR}}$, quasiparticle magnetic noise will incoherently drive NV spin transitions as shown in Fig. 2C, and the relaxation rate is given in terms of the amplitude of magnetic field fluctuations(*56*). Invoking an intuitive physical picture, when the NV sensor is scanned across physical edges of the 45° twisted BSCCO device, quasiparticles occupied in edge channels will drive spin depolarization along with potential contributions from the bulk state, leading to enhanced NV relaxation rates.

As a first step in gaining confidence in our method, we probe NV spin relaxation at two representative lateral sample positions: on and off a physical sample edge of the 45° twisted BSCCO device. Figure 2D presents the two corresponding NV spin relaxation spectra recorded at 50 K. One can see that the NV center shows a clearly accelerated spin relaxation from the initially prepared $m_s = 0$ (where $m_s$ denotes the magnetic quantum number) to an unpolarized spin state when positioned above the edge of the device, suggesting enhanced quasiparticle magnetic noise emanating from edge states. By measuring spin-dependent NV photoluminescence (PL) as a function of the delay time and fitting it to a theoretical model(*37*), the occupation probabilities of NV spin states can be obtained, allowing for quantitative extraction of NV spin relaxation rate Γ [see section S3 of (*40*) for details].

Next, we present one-dimensional (1D) scanning-probe NV relaxometry results to show direct evidence of topological edge modes in 45° twisted BSCCO. Figure 2F plots three linecuts of NV spin relaxation rate Γ measured at 50 K along the red, blue, and black dashed lines shown in the optical image (Fig. 2E). Notably, Γ shows a constant value over the bulk sample area while two peak signatures show up at the corresponding edge positions (red and blue points). The measured spin relaxation rate exhibits an inverse square dependence on the shortest distance between the NV center and sample edges as shown by a zoomed-in view in Fig. 2G, in agreement with our theoretical prediction [see section S4 of (*40*) for details]. It is instructive to note that critical fluctuations in BSCCO have a negligible effect on our results shown here as the measurement temperature (50 K) is far away from the $T_c$(*55*). Little variation of Γ is observed when the NV sensor is scanned over pristine BSCCO flakes and the substrate as shown by black points in Fig. 2F.

**Evaluation of topological band gap and topological transition temperature in twisted BSCCO**

For 45° twisted BSCCO, the gap at nodal points is expected to be a mixture of $d_{x^2-y^2}$ and $id_{xy}$ and the latter is responsible for gapping the $d_{x^2-y^2}$ nodal points to establish the chiral edge order(*7*). Considering that the interlayer Cooper tunneling strength is much smaller than that of the intralayer tunneling, only a small portion of $d_{xy}$ will be introduced in a real material environment(*7*). Thus, we use $d_{x^2-y^2} \pm i\epsilon d_{xy}$ to describe the order parameter of 45° twisted BSCCO, where $0 \leq \epsilon \leq 1$ defines the amplitude of the $d_{xy}$ part. Figure 3A plots three characteristic superconducting order parameter structures of 45° twisted BSCCO with different values of $\epsilon$. When $\epsilon = 0$, the two BSCCO flakes are basically decoupled and behave as two independent *d*-wave superconductors showing gapless nodal points. In the limit of $\epsilon = 1$, twisted BSCCO features an isotropic quasiparticle spectrum with a uniform gap. In a more realistic physical picture of $0 \leq \epsilon \leq 1$, 45° twisted BSCCO exhibits a maximum gap Δ that determines its superconducting phase transition while a gap of $\epsilon\Delta$ opened at the nodal provides topological protection to the edge modes.

Next, we present temperature dependent variations of edge-mode-induced NV spin relaxation rate to evaluate the magnitude of topological gap $\epsilon\Delta$ formed in twisted BSCCO. Figure 3B plots a series of 1D spectra of NV spin relaxation rate measured across an edge of the 45° twisted BSCCO device at different temperatures. Here, we have subtracted background of the bulk contribution to obtain $\Delta\Gamma$ [see section S4 of (*40*) for details], which highlights the edge-mode-induced NV spin relaxation. One can see that $\Delta\Gamma$ shows distinct variation trends in the surveyed temperature range from 5 K to 110 K, which is correlated to the temperature dependent topological band gap and quasiparticle population in edge states of twisted BSCCO as summarized in Fig. 3C. In the low temperature regime where $T \leq \epsilon\Delta$, $\Delta\Gamma$ increases linearly with $T$ as quasiparticles continue to populate available states in the edge channels. Thereafter, $\Delta\Gamma$ appears to saturate to a nearly constant value in the intermediate temperature regime. When approaching the $T_c$ ($T > 50$ K), the topological gap starts to decrease owing to reduction of the parental superconducting gap $\Delta$. Even though the pair of edge modes disperse chirally, they carry opposite currents. Thus, scattering between the edge modes, due to interaction with bulk quasiparticles, or fluctuations of the bulk superconducting order, leads to relaxation of the current and broadening of the associated spectral function, resulting in decrease of $\Delta\Gamma$ down to zero around the $T_c$. We have developed a detailed theoretical model to capture the observed temperature dependent variations of $\Delta\Gamma$ [see section S5 of (*40*) for details]:

$$\Delta\Gamma \sim \frac{\gamma_{\mathrm{NV}}^2 {\mu_0}^2 g_{\mathrm{s}} g_{\mathrm{c}} N_{\mathrm{L}}}{(4\pi)^2} \int_{-\infty}^{\infty} \frac{C_{JJ}(k, f_{\mathrm{NV}})}{d_{\mathrm{NV}}^2} f(k, z_{\mathrm{NV}}) dk \quad (1)$$

Where $\gamma_{\mathrm{NV}}$ is the gyromagnetic ratio, $g_s = 2$ is the spin degeneracy, $g_c = 2$ is the Chern number of the topological states in twisted BSCCO(*7*), $N_L$ is the number of effective BSCCO layers, and $C_{JJ}$ is the current spectral density incorporating temperature $T$ and the topological band gap $\epsilon\Delta$ [see section S6 of (*40*) for details], $k$ is wavevector of quasiparticles, $d_{\mathrm{NV}}$ is the shortest distance between the NV center and the sample edge, $z_{\mathrm{NV}}$ is the vertical NV-to-sample distance, and $f(k, z_{\mathrm{NV}})$ is a transfer function describing magnetic fields generated at the NV site [see section S4 of (*40*) for details]. Our experimental results can be fitted well to the theoretical model, from which $\epsilon\Delta$ is obtained to be ~4 meV at 0 K, in qualitative agreement with previous theoretical calculations [see sections S5 and S6 of (*40*) for details](*7*). It is worth mentioning that NV spin relaxation rate measured in a pristine BSCCO sample area shows a characteristic $\sim T^2$ dependence, consistent with the model of quasiparticle-induced magnetic noise for a nodal *d*-wave superconductor [see section S7 of (*40*) for details](*55*).

To investigate twist-controlled topology, we further perform NV spin relaxometry measurements on BSCCO devices with different twisted angles as presented in Fig. 3C [see section S8 of (*40*) for details]. Overall, $\Delta\Gamma(T)$ measured on 43° twisted BSCCO follows a similar temperature dependence of 45° samples in the low temperature range ($T < 20$ K). As we discussed above, the low temperature increase of $\Delta\Gamma(T)$ is governed by the zero-temperature topological gap, suggesting that $\epsilon\Delta$ is insensitive in this twisting angle range. We notice that $\Delta\Gamma(T)$ measured in the 43° sample vanishes at $\tilde{T}_c = 70$ K, clearly lower than the bulk $T_c = 86$ K. This is a sharp contrast for the 45° twisted samples where $\tilde{T}_c \approx T_c$ is observed. For $\theta = 37°$, the low temperature increase behavior of $\Delta\Gamma(T)$ persists, however, the measured topological transition temperature $\tilde{T}_c$ dramatically decreases to ~30 K. When $\theta = 37°$ and 24°, the measured $\Delta\Gamma$ becomes vanishingly small over the entire temperature range surveyed, indicating absence of edge modes.

For more quantitative analysis, we employ Eq. (1) against to the experimentally measured $\Delta\Gamma(T)$ for samples with different twisting angles $\theta$ and obtain both $\epsilon\Delta(\theta)$ and $\tilde{T}_c(\theta)$. Figure 3D summarizes the obtained topological bandgap $\epsilon\Delta$ as a function of $\theta$ for twisted BSCCO, where a critical angle $\theta_c$ between 37° and 40° is observed below which $\epsilon\Delta$ remains to be zero. Above the $\theta_c$, $\epsilon\Delta$ basically shows a constant value of ~4 meV (at 0 K), in agreement with our theoretical calculations [see section S6 of (*40*) for details]. Twist angle dependences of $\tilde{T}_c(\theta)$ and $T_c(\theta)$ are presented in Fig. 3E. One can see that the topological transition temperature $\tilde{T}_c$ monotonically increases when $\theta > \theta_c$ and eventually rises to the value of $T_c$ when $\theta = 45°$. In contrast, superconducting transition temperature $T_c$ characterized by the maximum band gap Δ in twisted BSCCO shows a constant value (~85 K) irrespective of variations of $\theta$. The observed experimental behaviors are in agreement with theoretical expectation of $\tilde{T}_c(\theta)$, below which two degenerated Ginzberg-Landau (GL) free energy minima are developed. As $\theta$ moves away from the optimal twist angle (45°), the GL free energy minima, establishing topological edge state of twisted BSCCO, become shallower(*7*) and the topological band gap closes at $\tilde{T}_c < T_c$.

**Field-induced topology in twisted BSCCO**

Lastly, we explore experimental signatures of magnetic field-induced topology in 45° twisted BSCCO. With application of a moderate in-plane magnetic field, Josephson vortices are predicted to form along the thickness direction of 45° twisted BSCCO, creating a periodic lattice of alternating topological domains of $d_{x^2-y^2} \pm id_{xy}$ as illustrated in Fig. 4A(*27*, *34*). The two neighboring $d_{x^2-y^2} \pm id_{xy}$ domains show a phase difference of 180° and support edge currents with opposite chirality. Domain walls naturally define chiral edge states between neighboring topological domains with opposite Chern numbers. Here, we leverage scanning NV gradiometry(*43*) to detect Oersted fields generated by chiral edge currents in 45° twisted BSCCO [see sections S9 and S10 of (*40*) for details].

Figures 4B and 4C present static magnetic field $B_c$ maps of a surveyed sample area under different external field conditions. Note that $B_c$ is measured along the NV spin direction and the sample area investigated is away from physical edges of twisted BSCCO. When an in-plane magnetic field $B_y$ = 20 G is applied, periodic field domains with alternatingly opposite polarity emerge along the transverse direction (in relative to $B_y$) in 45° twisted BSCCO as shown in Fig. 4B. When direction of the external magnetic field is reversed ($B_y$ = −20 G), shape of the field pattern basically remains unchanged while polarity of individual domains switches accordingly (Fig. 4C), which reminisces the signature of theoretically predicted Josephson vortex lattice(*27*, *34*). When $B_y$ = 0, the periodic magnetic field domains disappear [see section S11 of (*40*) for details], consistent with the absence of field-induced topological $d_{x^2-y^2} \pm id_{xy}$ domains in twisted nodal superconductors. Invoking the Biot-Savart law, the corresponding transverse and longitudinal 2D chiral electric current density $J_x$ and $J_y$ as well as the current flow pattern (when $B_y$ = −20 G) can be reconstructed as shown in Figs. 4D-4F(*43*, *57*). Theoretically, we can expect that the longitudinal electric current density $J_y$ at domain walls due to contributions from two neighboring domains is approximately twice as large as the transverse component $J_x$. Figure 4G presents linecuts of $J_x$ and $J_y$ measured along the red and green dashed lines shown in Figs. 4, D and E. The peak values of $J_y$ and $J_x$ are determined to be ~2.0 A/m and ~0.8 A/m, respectively, consistent with the physical picture discussed above. Magnetic hysteresis and temperature dependence of the periodic field pattern are also investigated as presented in section S11 of (*40*).

**Conclusions**

In this study, we report scanning-probe quantum sensing of twisted vdW cuprate superconductors. Chiral edge modes in twisted BSCCO are visualized using NV spin relaxometry method. By investigating temperature dependent variations of NV relaxation rate, we present an experimental approach to evaluate the magnitude of topological band gap in twisted BSCCO. We also observe experimental signatures of field-induced chiral topological domains and edge modes in 45° twisted BSCCO. Our results are rationalized well by theoretical calculations, bringing insights into the interplay between twist lattice engineering, unconventional superconductivity, and topology in a highly expected topological superconductor platform. The scanning-probe quantum sensing microscopy presented also opens new pathways for probing nanoscale topological and correlated phases of a broad family of moiré quantum matter(*19*, *58–60*).

**Data availability**. All data supporting the findings of this study are available from the corresponding authors on reasonable request.

**Acknowledgements**. The authors are grateful to Pavel Volko and Itamar Kimchi for helpful discussions. The quantum sensing measurements are based upon work supported by the Air Force Office of Scientific Research under award No. FA9550-25-1-0082. Development of experimental hardware for cryogenic quantum microscopy was supported by the U.S. Department of Energy (DOE), Office of Science, Basic Energy Sciences (BES), under award No. DE-SC0024870. Material/device characterizations at Georgia Tech were supported by the Office of Naval Research (ONR) under grant No. N00014-23-1-2146. H. W. acknowledges the support from the U.S. National Science Foundation under award No. ECCS-2525800. P. K. acknowledges support from AFOSR (FA9550-25-1-0019) for data analysis, and X. C. acknowledges support from DOE/BES/QIS (DE-SC0026183) for developing sample transfer methods. The theoretical calculations at Argonne National Laboratory were supported by the US Department of Energy, Office of Science, Basic Energy Sciences, Material Science & Technology Division (FWP number PRJ1000084, K. A.). The crystal growth at Brookhaven National Laboratory was supported by the US Department of Energy, Office of Basic Energy Sciences, Materials Sciences and Engineering Division, under contract no. DOE-SC0012704 (G. G.). K. W. and T. T. acknowledge support from the JSPS KAKENHI (Grant Numbers 21H05233 and 23H02052), the CREST (JPMJCR24A5), JST and World Premier International Research Center Initiative (WPI), MEXT, Japan.

**Author contributions**.
S. L. performed the NV measurements and analyzed the data with R. S., L. Z., Y. Z. and H. W. X. C. and P. K. provided the twisted BSCCO devices. K. A. performed theoretical calculations on edge-mode-induced magnetic noise in twisted BSCCO. G. G. provided bulk BSCCO crystals. K. W. and T. T. provided bulk hBN crystals. H. W., K. A., X. C., P. K. and C. R. D. discussed the results and contributed to writing the manuscript. C. R. D. supervised this project.

**Competing interests**
The authors declare no competing interests.

## References:

1. Y. Cao, V. Fatemi, S. Fang, K. Watanabe, T. Taniguchi, E. Kaxiras, P. Jarillo-Herrero, Unconventional superconductivity in magic-angle graphene superlattices. *Nature* **556**, 43–50 (2018).
2. L. Balents, C. R. Dean, D. K. Efetov, A. F. Young, Superconductivity and strong correlations in moiré flat bands. *Nat. Phys.* **16**, 725–733 (2020).
3. D. M. Kennes, M. Claassen, L. Xian, A. Georges, A. J. Millis, J. Hone, C. R. Dean, D. N. Basov, A. N. Pasupathy, A. Rubio, Moiré heterostructures as a condensed-matter quantum simulator. *Nat. Phys.* **17**, 155–163 (2021).
4. K. S. Burch, D. Mandrus, J.-G. Park, Magnetism in two-dimensional van der Waals materials. *Nature* **563**, 47–52 (2018).
5. A. K. Geim, I. V. Grigorieva, Van der Waals heterostructures. *Nature* **499**, 419–425 (2013).
6. S. Y. F. Zhao, X. Cui, P. A. Volkov, H. Yoo, S. Lee, J. A. Gardener, A. J. Akey, R. Engelke, Y. Ronen, R. Zhong, G. Gu, S. Plugge, T. Tummuru, M. Kim, M. Franz, J. H. Pixley, N. Poccia, P. Kim, Time-reversal symmetry breaking superconductivity between twisted cuprate superconductors. *Science* **382**, 1422–1427 (2023).
7. O. Can, T. Tummuru, R. P. Day, I. Elfimov, A. Damascelli, M. Franz, High-temperature topological superconductivity in twisted double-layer copper oxides. *Nat. Phys.* **17**, 519–524 (2021).
8. Y. Xia, Z. Han, K. Watanabe, T. Taniguchi, J. Shan, K. F. Mak, Superconductivity in twisted bilayer $WSe_2$. *Nature* **637**, 833–838 (2025).
9. Y. Guo, J. Pack, J. Swann, L. Holtzman, M. Cothrine, K. Watanabe, T. Taniguchi, D. G. Mandrus, K. Barmak, J. Hone, A. J. Millis, A. Pasupathy, C. R. Dean, Superconductivity in 5.0° twisted bilayer $WSe_2$. *Nature* **637**, 839–845 (2025).
10. Z. Wang, B. Xia, S. Paolini, Z.-J. Yan, P. Xiao, J. Song, V. Gowda, H. Rong, D. Xiao, X. Xu, W. Wu, Z. Wang, C.-Z. Chang, Moiré engineering of Cooper-pair density modulation states. *Nature* **652**, 335–341 (2026).
11. Y. Xu, A. Ray, Y.-T. Shao, S. Jiang, K. Lee, D. Weber, J. E. Goldberger, K. Watanabe, T. Taniguchi, D. A. Muller, K. F. Mak, J. Shan, Coexisting ferromagnetic–antiferromagnetic state in twisted bilayer $CrI_3$. *Nat. Nanotechnol.* **17**, 143–147 (2022).
12. T. Song, Q.-C. Sun, E. Anderson, C. Wang, J. Qian, T. Taniguchi, K. Watanabe, M. A. McGuire, R. Stöhr, D. Xiao, T. Cao, J. Wrachtrup, X. Xu, Direct visualization of magnetic domains and moiré magnetism in twisted 2D magnets. *Science* **374**, 1140–1144 (2021).
13. H. Xie, X. Luo, G. Ye, Z. Ye, H. Ge, S. H. Sung, E. Rennich, S. Yan, Y. Fu, S. Tian, H. Lei, R. Hovden, K. Sun, R. He, L. Zhao, Twist engineering of the two-dimensional magnetism in double bilayer chromium triiodide homostructures. *Nat. Phys.* **18**, 30-36 (2022).
14. G. Cheng, M. M. Rahman, A. L. Allcca, X. Liu, L. Liu, L. Fu, Y. Zhu, Z. Mao, K. Watanabe, P. Upadhyaya, Y. P. Chen, Electrically tunable moiré magnetism in twisted double bilayers of chromium triiodide. *Nat. Electron.* **6**, 434–442 (2023).
15. S. Li, Z. Sun, N. J. McLaughlin, A. Sharmin, N. Agarwal, M. Huang, S. H. Sung, H. Lu, S. Yan, H. Lei, R. Hovden, H. Wang, H. Chen, L. Zhao, C. R. Du, Observation of stacking engineered magnetic phase transitions within moiré supercells of twisted van der Waals magnets. *Nat. Commun.* **15**, 5712 (2024).
16. K. Tran, G. Moody, F. Wu, X. Lu, J. Choi, K. Kim, A. Rai, D. A. Sanchez, J. Quan, A. Singh, J. Embley, A. Zepeda, M. Campbell, T. Autry, T. Taniguchi, K. Watanabe, N. Lu, S. K. Banerjee,

K. L. Silverman, S. Kim, E. Tutuc, L. Yang, A. H. MacDonald, X. Li, Evidence for moiré excitons in van der Waals heterostructures. *Nature* **567**, 71–75 (2019).

17. C. Jin, E. C. Regan, A. Yan, M. Iqbal Bakti Utama, D. Wang, S. Zhao, Y. Qin, S. Yang, Z. Zheng, S. Shi, K. Watanabe, T. Taniguchi, S. Tongay, A. Zettl, F. Wang, Observation of moiré excitons in $WSe_2/WS_2$ heterostructure superlattices. *Nature* **567**, 76–80 (2019).
18. K. L. Seyler, P. Rivera, H. Yu, N. P. Wilson, E. L. Ray, D. G. Mandrus, J. Yan, W. Yao, X. Xu, Signatures of moiré-trapped valley excitons in $MoSe_2/WSe_2$ heterobilayers. *Nature* **567**, 66–70 (2019).
19. L. Ju, A. H. MacDonald, K. F. Mak, J. Shan, X. Xu, The fractional quantum anomalous Hall effect. *Nat. Rev. Mater.* **9**, 455–459 (2024).
20. Y. Xie, A. T. Pierce, J. M. Park, D. E. Parker, E. Khalaf, P. Ledwith, Y. Cao, S. H. Lee, S. Chen, P. R. Forrester, K. Watanabe, T. Taniguchi, A. Vishwanath, P. Jarillo-Herrero, A. Yacoby, Fractional Chern insulators in magic-angle twisted bilayer graphene. *Nature* **600**, 439–443 (2021).
21. B. A. Foutty, C. R. Kometter, T. Devakul, A. P. Reddy, K. Watanabe, T. Taniguchi, L. Fu, B. E. Feldman, Mapping twist-tuned multiband topology in bilayer $WSe_2$. *Science* **384**, 343–347 (2024).
22. Y. Yu, L. Ma, P. Cai, R. Zhong, C. Ye, J. Shen, G. D. Gu, X. H. Chen, Y. Zhang, High-temperature superconductivity in monolayer $Bi_2Sr_2CaCu_2O_{8+\delta}$. *Nature* **575**, 156–163 (2019).
23. A. von Hoegen, T. Tai, C. J. Allington, M. Yeung, J. Pettine, M. H. Michael, E. Viñas Boström, X. Cui, K. Torres, A. E. Kossak, B. Lee, G. S. D. Beach, G. D. Gu, A. Rubio, P. Kim, N. Gedik, Imaging a terahertz superfluid plasmon in a two-dimensional superconductor. *Nature* **650**, 869–874 (2026).
24. Y. He, M. Hashimoto, D. Song, S.-D. Chen, J. He, I. M. Vishik, B. Moritz, D.-H. Lee, N. Nagaosa, J. Zaanen, T. P. Devereaux, Y. Yoshida, H. Eisaki, D. H. Lu, Z.-X. Shen, Rapid change of superconductivity and electron-phonon coupling through critical doping in Bi-2212. *Science* **362**, 62–65 (2018).
25. H. Ding, T. Yokoya, J. C. Campuzano, T. Takahashi, M. Randeria, M. R. Norman, T. Mochiku, K. Kadowaki, J. Giapintzakis, Spectroscopic evidence for a pseudogap in the normal state of underdoped high-$T_c$ superconductors. *Nature* **382**, 51–54 (1996).
26. N. Avraham, B. Khaykovich, Y. Myasoedov, M. Rappaport, H. Shtrikman, D. E. Feldman, T. Tamegai, P. H. Kes, M. Li, M. Konczykowski, K. van der Beek, E. Zeldov, "Inverse" melting of a vortex lattice. *Nature* **411**, 451–454 (2001).
27. P. A. Volkov, J. H. Wilson, K. P. Lucht, J. H. Pixley, Current- and field-induced topology in twisted nodal superconductors. *Phys. Rev. Lett.* **130**, 186001 (2023).
28. C. Nayak, S. H. Simon, A. Stern, M. Freedman, S. Das Sarma, Non-Abelian anyons and topological quantum computation. *Rev. Mod. Phys.* **80**, 1083–1159 (2008).
29. S. Qi, J. Ge, C. Ji, Y. Ai, G. Ma, Z. Wang, Z. Cui, Y. Liu, Z. Wang, J. Wang, High-temperature field-free superconducting diode effect in high-$T_c$ cuprates. *Nat. Commun.* **16**, 531 (2025).
30. S. Ghosh, V. Patil, A. Basu, Kuldeep, A. Dutta, D. A. Jangade, R. Kulkarni, A. Thamizhavel, J. F. Steiner, F. von Oppen, M. M. Deshmukh, High-temperature Josephson diode. *Nat. Mater.* **23**, 612–618 (2024).
31. H. Wang, Y. Zhu, Z. Bai, Z. Lyu, J. Yang, L. Zhao, X. J. Zhou, Q.-K. Xue, D. Zhang, Quantum superconducting diode effect with perfect efficiency above liquid-nitrogen temperature. *Nat. Phys.* **22**, 47–53 (2026).

32. J. Lee, W. Lee, G.-Y. Kim, Y.-B. Choi, J. Park, S. Jang, G. Gu, S.-Y. Choi, G. Y. Cho, G.-H. Lee, H.-J. Lee, Twisted van der Waals Josephson junction based on a high-$T_c$ superconductor. *Nano Lett.* **21**, 10469–10477 (2021).
33. T. Confalone, F. Lo Sardo, Y. Lee, S. Shokri, G. Serpico, A. Coppo, L. Chirolli, V. M. Vinokur, V. Brosco, U. Vool, D. Montemurro, F. Tafuri, K. Nielsch, G. Haider, N. Poccia, Cuprate twistronics for quantum hardware. *Adv. Quantum Technol.* **8**, 2500203 (2025).
34. V. Pathak, O. Can, M. Franz, Edge currents as probe of topology in twisted cuprate bilayers. *Phys. Rev. B* **110**, 014506 (2024).
35. J. H. Pixley, P. A. Volkov, Twisted nodal superconductors. *Annu. Rev. Condes. Matter. Rev.* **17**, 183-205 (2026).
36. S. Kolkowitz, A. Safira, A. A. High, R. C. Devlin, S. Choi, Q. P. Unterreithmeier, D. Patterson, A. S. Zibrov, V. E. Manucharyan, H. Park, M. D. Lukin, Probing Johnson noise and ballistic transport in normal metals with a single-spin qubit. *Science* **347**, 1129–1132 (2015).
37. A. Ariyaratne, D. Bluvstein, B. A. Myers, A. C. B. Jayich, Nanoscale electrical conductivity imaging using a nitrogen-vacancy center in diamond. *Nat. Commun.* **9**, 2406 (2018).
38. A. Finco, A. Haykal, R. Tanos, F. Fabre, S. Chouaieb, W. Akhtar, I. Robert-Philip, W. Legrand, F. Ajejas, K. Bouzehouane, N. Reyren, T. Devolder, J.-P. Adam, J.-V. Kim, V. Cros, V. Jacques, Imaging non-collinear antiferromagnetic textures via single spin relaxometry. *Nat. Commun.* **12**, 767 (2021).
39. B. A. McCullian, A. M. Thabt, B. A. Gray, A. L. Melendez, M. S. Wolf, V. L. Safonov, D. V. Pelekhov, V. P. Bhallamudi, M. R. Page, P. C. Hammel, Broadband multi-magnon relaxometry using a quantum spin sensor for high frequency ferromagnetic dynamics sensing. *Nat. Commun.* **11**, 5229 (2020).
40. See Supplemental Material for details.
41. L. Thiel, D. Rohner, M. Ganzhorn, P. Appel, E. Neu, B. Müller, R. Kleiner, D. Koelle, P. Maletinsky, Quantitative nanoscale vortex imaging using a cryogenic quantum magnetometer. *Nat. Nanotechnol.* **11**, 677–681 (2016).
42. R. Monge, T. Delord, N. V. Proscia, Z. Shotan, H. Jayakumar, J. Henshaw, P. R. Zangara, A. Lozovoi, D. Pagliero, P. D. Esquinazi, T. An, I. Sodemann, V. M. Menon, C. A. Meriles, Spin dynamics of a solid-state qubit in proximity to a superconductor. *Nano Lett.* **23**, 422–428 (2023).
43. W. S. Huxter, M. L. Palm, M. L. Davis, P. Welter, C.-H. Lambert, M. Trassin, C. L. Degen, Scanning gradiometry with a single spin quantum magnetometer. *Nat. Commun.* **13**, 3761 (2022).
44. A. K. C. Tan, H. Jani, M. Högen, L. Stefan, C. Castelnovo, D. Braund, A. Geim, A. Mechnich, M. S. G. Feuer, H. S. Knowles, A. Ariando, P. G. Radaelli, M. Atatüre, Revealing emergent magnetic charge in an antiferromagnet with diamond quantum magnetometry. *Nat. Mater.* **23**, 205–211 (2024).
45. Q. Guo, A. D'Addario, Y. Cheng, J. Kline, I. Gray, H. F. H. Cheung, F. Yang, K. C. Nowack, G. D. Fuchs, Current-induced switching of thin film $\alpha$-$Fe_2O_3$ devices imaged using a scanning single-spin microscope. *Phys. Rev. Mater.* **7**, 064402 (2023).
46. S. Jayaram, M. Lenger, D. Zhao, L. Pupim, K. Watanabe, T. Taniguchi, R. Peng, M. Scheffler, R. Stöhr, M. S. Scheurer, J. Smet, J. Wrachtrup, Probing vortex dynamics in 2D superconductors with scanning quantum microscope. *Phys. Rev. Lett.* **135**, 126001 (2025).
47. S. Hsieh, P. Bhattacharyya, C. Zu, T. Mittiga, T. J. Smart, F. Machado, B. Kobrin, T. O. Höhn, N. Z. Rui, M. Kamrani, S. Chatterjee, S. Choi, M. Zaletel, V. V. Struzhkin, J. E. Moore, V. I.

Levitas, R. Jeanloz, N. Y. Yao, Imaging stress and magnetism at high pressures using a nanoscale quantum sensor. *Science* **366**, 1349–1354 (2019).
48. K. Y. Yip, K. O. Ho, K. Y. Yu, Y. Chen, W. Zhang, S. Kasahara, Y. Mizukami, T. Shibauchi, Y. Matsuda, S. K. Goh, S. Yang, Measuring magnetic field texture in correlated electron systems under extreme conditions. *Science* **366**, 1355–1359 (2019).
49. M. Lesik, T. Plisson, L. Toraille, J. Renaud, F. Occelli, M. Schmidt, O. Salord, A. Delobbe, T. Debuisschert, L. Rondin, P. Loubeyre, J.-F. Roch, Magnetic measurements on micrometer-sized samples under high pressure using designed NV centers. *Science* **366**, 1359–1362 (2019).
50. J. Rovny, Z. Yuan, M. Fitzpatrick, A. I. Abdalla, L. Futamura, C. Fox, M. C. Cambria, S. Kolkowitz, N. P. de Leon, Nanoscale covariance magnetometry with diamond quantum sensors. *Science* **378**, 1301–1305 (2022).
51. M. Borst, P. H. Vree, A. Lowther, A. Teepe, S. Kurdi, I. Bertelli, B. G. Simon, Y. M. Blanter, T. van der Sar, Observation and control of hybrid spin-wave–Meissner-current transport modes. *Science* **382**, 430–434 (2023).
52. Y. Schlussel, T. Lenz, D. Rohner, Y. Bar-Haim, L. Bougas, D. Groswasser, M. Kieschnick, E. Rozenberg, L. Thiel, A. Waxman, J. Meijer, P. Maletinsky, D. Budker, R. Folman, Wide-field imaging of superconductor vortices with electron spins in diamond. *Phys. Rev. Appl.* **10**, 034032 (2018).
53. Y. Xu, Y. Yu, Y. Y. Hui, Y. Su, J. Cheng, H.-C. Chang, Y. Zhang, Y. R. Shen, C. Tian, Mapping dynamical magnetic responses of ultrathin micron-size superconducting films using nitrogen-vacancy centers in diamond. *Nano Lett.* **19**, 5697–5702 (2019).
54. D. D. Awschalom, R. Hanson, J. Wrachtrup, B. B. Zhou, Quantum technologies with optically interfaced solid-state spins. *Nat. Photon.* **12**, 516–527 (2018).
55. Z. Liu, R. Gong, J. Kim, O. K. Diessel, Q. Xu, Z. Rehfuss, X. Du, G. He, A. Singh, Y. S. Eo, E. A. Henriksen, G. D. Gu, N. Y. Yao, F. Machado, S. Ran, S. Chatterjee, C. Zu, Quantum noise spectroscopy of superconducting dynamics in thin film $Bi_2Sr_2CaCu_2O_{8+\delta}$. Preprint at https://doi.org/10.48550/arXiv.2502.04439 (2025).
56. S. Li, S. P. Kelly, J. Zhou, H. Lu, Y. Tserkovnyak, H. Wang, C. R. Du, Nanoscale observation and control of quasiparticle induced magnetic noise in a superconducting resonator. *Phys. Rev. Lett.* **136**, 076004 (2026).
57. S. Midha, M. Parashar, A. Bathla, D. A. Broadway, J.-P. Tetienne, K. Saha, Optimized current-density reconstruction from wide-field quantum diamond magnetic field maps. *Phys. Rev. Appl.* **22**, 014015 (2024).
58. J. F. Rodriguez-Nieva, K. Agarwal, T. Giamarchi, B. I. Halperin, M. D. Lukin, E. Demler, Probing one-dimensional systems via noise magnetometry with single spin qubits. *Phys. Rev. B* **98**, 195433 (2018).
59. T. Han, Z. Lu, Y. Yao, L. Shi, J. Yang, J. Seo, S. Ye, Z. Wu, M. Zhou, H. Liu, G. Shi, Z. Hua, K. Watanabe, T. Taniguchi, P. Xiong, L. Fu, L. Ju, Signatures of chiral superconductivity in rhombohedral graphene *Nature* **643**, 654–661 (2025).
60. Y.-H. Li, R. Cheng, Moiré magnons in twisted bilayer magnets with collinear order. *Phys. Rev. B* **102**, 094404 (2020).
61. B. J. Shields, Q. P. Unterreithmeier, N. P. de Leon, H. Park, M. D. Lukin, Efficient readout of a single spin state in diamond via spin-to-charge conversion. *Phys. Rev. Lett.* **114**, 136402 (2015).
62. W. Huang, E. Taylor, C. Kallin, Vanishing edge currents in non-*p*-wave topological chiral superconductors. *Phys. Rev. B* **90**, 224519 (2014).

63. M. Stone, R. Roy, Edge modes, edge currents, and gauge invariance in $p_x+ip_y$ superfluids and superconductors. *Phys. Rev. B* **69**, 184511 (2004).
64. S.-I. Suzuki, Y. Asano, Spontaneous edge current in a small chiral superconductor with a rough surface. *Phys. Rev. B* **94**, 155302 (2016).
65. K. Agarwal, R. Schmidt, B. Halperin, V. Oganesyan, G. Zaránd, M. D. Lukin, E. Demler, Magnetic noise spectroscopy as a probe of local electronic correlations in two-dimensional systems. *Phys. Rev. B* **95**, 155107 (2017).
66. Midha, S., Singh, R., Gharavi, K., Baugh, J. & Muralidharan, B. On the microscopics of proximity effects in one-dimensional superconducting hybrid systems. Preprint at https://doi.org/10.48550/arXiv.2411.12733 (2024).
67. G. Rai, S. Haas, A. Jagannathan, Superconducting proximity effect and order parameter fluctuations in disordered and quasiperiodic systems. *Phys. Rev. B* **102**, 134211 (2020).

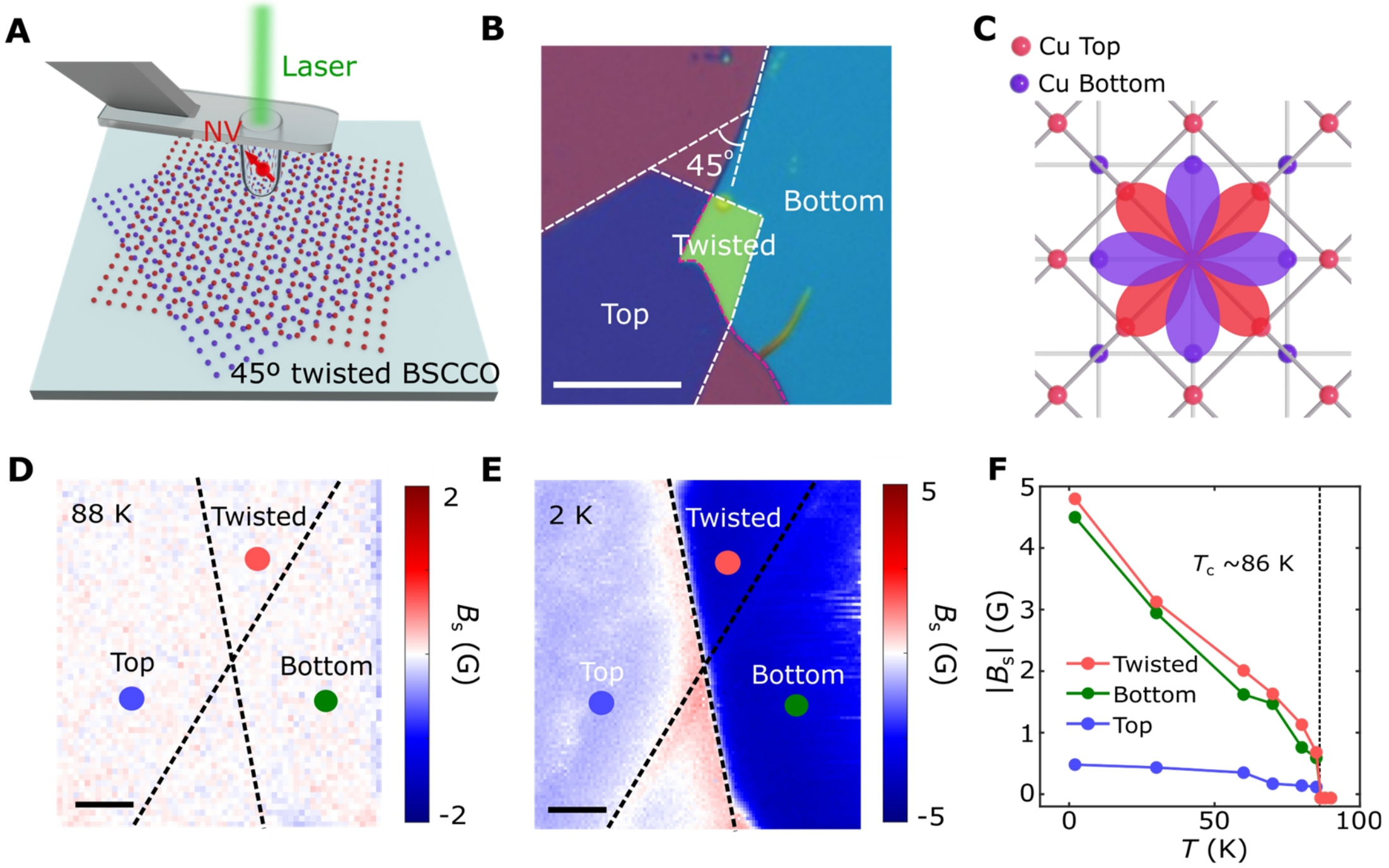

**Fig. 1. Scanning-probe quantum sensing of twisted BSCCO.** (**A**) Schematic illustration of scanning NV measurements of 45° twisted BSCCO. (**B**) Optical microscopy image of a prepared 45° twisted BSCCO device. Boundaries of the top and bottom BSCCO layers are outlined by dashed lines. The top, bottom, and twisted BSCCO sample areas are denoted. The scale bar is 20 μm. (**C**) Hybrid superconducting order parameter $d_{x^2-y^2} \pm id_{xy}$ of 45° twisted BSCCO. (**D** and **E**) Scanning NV imaging of supercurrent-induced Meissner effect in BSCCO at 88 K and 2 K. The scale bar is 1 μm. (**F**) Temperature dependence of supercurrent-induced magnetic field $|B_s|$ measured at three local sites in the twisted, top, and bottom BSCCO areas. Red, blue, and green points in Figs. 1D-1E denote the three sample positions surveyed for local NV sensing measurements.

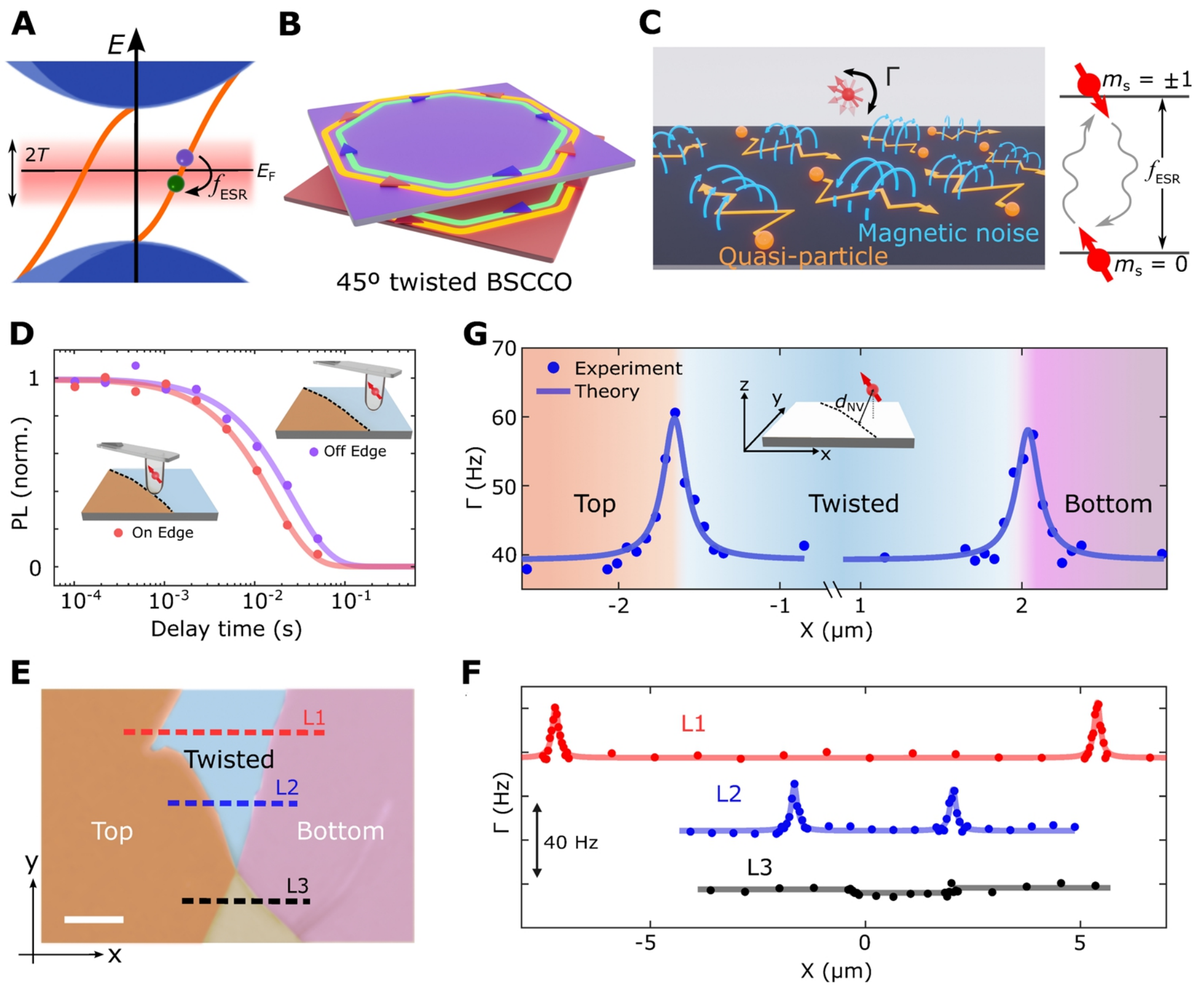

**Fig. 2. Visualizing gapless edge modes in 45° twisted BSCCO.** (**A**) Topological band gap and gapless chiral edge states formed at nodal points of 45° twisted BSCCO. The red color shaded area represents chiral edge states populated with quasiparticles. $E_F$ denotes the Fermi energy and $T$ is temperature. (**B**) Schematic of two edge modes carrying counterpropagating currents in top and bottom layers of 45° twisted BSCCO. (**C**) Thermally induced random motions of quasiparticles in twisted BSCCO generate fluctuating magnetic fields at ESR frequency $f_{ESR}$ driving relaxation of a proximal NV center. (**D**) Two sets of NV spin relaxometry spectra recorded when the NV sensor is positioned on and off a physical edge of a 45° twisted BSCCO device. NV photoluminescence (PL) is measured as a function of the delay time. (**E**) False-colored microscopy image denoting the top (orange), bottom (purple), and twisted (blue) BSCCO sample areas. The scale bar is 5 μm. (**F**) NV spin relaxation rate Γ measured along the red ($L_1$), blue ($L_2$), and black ($L_3$) dashed lines shown in Fig. 2E. Scanning directions of the NV sensor are defined as *x*-axis in the local coordinate frame. Solid lines are fittings of experimental results (dots) to our theoretical model. (**G**) A zoomed-in view of 1D variation of Γ measured over physical edges of the 45° twisted BSCCO device. Orange, blue, and purple background colors highlight the top, 45° twisted, and bottom BSSCO sample regions. The vertical NV-to-sample distance is ~60 nm and measurement temperature is 50 K for all the NV sensing results presented in Fig. 2.

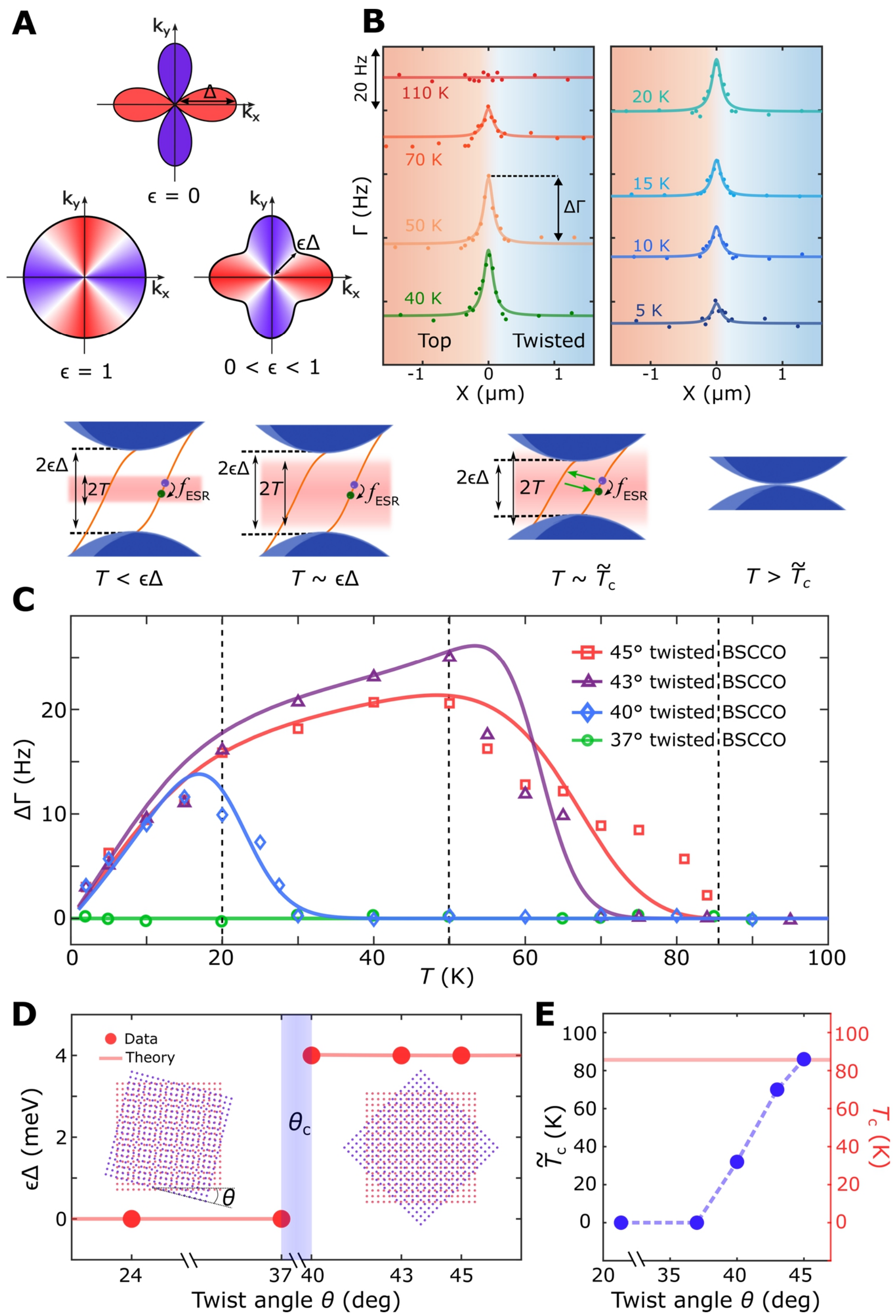

**Fig. 3. Variations of topological bandgap and topological transition temperatures in twisted BSCCO.** (**A**) Characteristic superconducting order parameter structures of 45° twisted BSCCO with different amplitude of the $d_{xy}$ part. (**B**) A series of 1D scanning NV relaxometry results measured across a physical edge of the 45° twisted BSCCO device at temperatures from 5 K to 110 K. Solid lines are fittings of experimental results (dots) to theoretical model. The origin of $x$-axis has been shifted to the position of device edge for visual clarity. (**C**) Temperature dependent

variations of edge-mode-induced NV spin relaxation rate $\Delta\Gamma$ for 45°, 43°, 40°, and 37° twisted BSCCO. Solid lines are fittings of experimental results (points) to our theory. Sketch of temperature dependent evolution of topological band gap opened at nodal points of 45° twisted BSCCO are depicted on top. Red shaded color area increasing with temperature represents the available edge states populated with quasiparticles. Green arrows in the case of $T \sim \tilde{T}_c$ illustrate scattering between the two edge modes. The vertical NV-to-sample distance is ~60 nm for NV measurement results presented. (**D**) Topological bandgap $\epsilon\Delta$ (0 K) at nodal points of twisted BSCCO as a function of the twist angle $\theta$. The shaded color area from 37° to 40° represents the range of experimentally determined critical twist angle $\theta_c$. (**E**) Twist angle dependence of topological and superconducting transition temperatures $\tilde{T}_c$ and $T_c$.

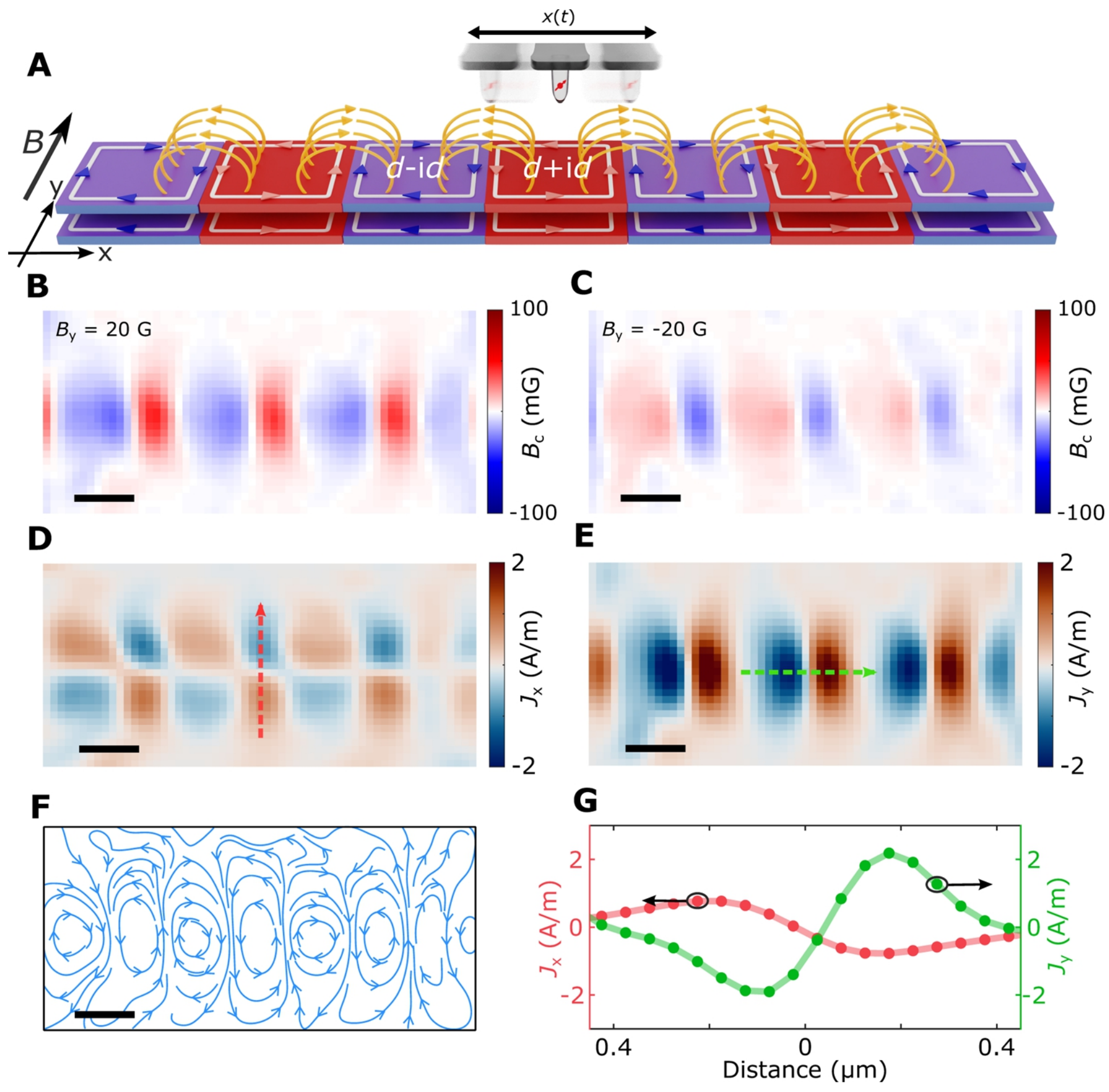

**Fig. 4. Field-induced topology in 45° twisted BSCCO.** (**A**) Schematic of scanning NV gradiometry to detect in-plane magnetic field-induced alternating topological domains of the $d_{x^2-y^2} \pm id_{xy}$ phase. (**B** and **C**) Static field maps of a surveyed sample area of 45° twisted BSCCO with application of an in-plane magnetic field $B_y = 20$ G (B) and $B_y = -20$ G (C). The measurement temperature is 2 K, and vertical NV-to-sample distance is ~80 nm. (**D**)-(**F**) Reconstructed 2D electric current density $J_x$ (D), $J_y$ (E), and current flow pattern (F) when $B_y = -20$ G. The scale bar is 400 nm in Figs. 4B-4F. (**G**) 1D linecuts of $J_x$ (red) and $J_y$ (green) measured along the red and green dashed arrows shown in Figs. 4D-4E.